\documentclass[12pt,a4paper]{article}

\usepackage[utf8]{inputenc}
\usepackage[margin=1in]{geometry}
\usepackage[english]{babel}
\usepackage{amsmath}
\usepackage{amsthm}
\newtheorem{theorem}{Theorem}
\newtheorem{lemma}{Lemma}
\usepackage{listings}
\usepackage{amsfonts}
\usepackage{amssymb}
\usepackage{authblk}
\usepackage{appendix}
\usepackage[linkcolor=black,colorlinks=true,citecolor=black,filecolor=black,urlcolor=black]{hyperref}

\author[1]{Petru\c{t} Cob\^arzan}
\affil[1]{Computer Science graduate student, ETH Z\"urich\\ 
\href{mailto:corvin-petrut.cobarzan@alumni.ethz.ch}{corvin-petrut.cobarzan@alumni.ethz.ch}}
\title{A new upper bound for Achlioptas processes}

\begin{document}

\maketitle

\begin{abstract}

We consider here on-line algorithms for Achlioptas processes. Given a initially empty graph $G$ on $n$ vertices, a random process that at each step selects independently and uniformly at random two edges from the set of non-edges is launched. We must choose one of the two edges and add it to the graph while discarding the other. The goal is to avoid the appearance of a connected component spanning $\Omega(n)$ vertices (called a giant component) for as many steps as possible.

Bohman and Frieze proved in 2001 that on-line Achlioptas processes cannot postpone the appearance of the giant for more that roughly $n$ steps whp. This upper bound got even lower in 2003 when the two above mentioned authors and Wormald proved that each on-line Achlioptas process creates a giant before step $0.964446n$ whp.

The purpose of this work is to determine a new upper bound. By using essentially the same methods used by Bohman, Frieze and Wormald in 2003 and some results of Spencer and Wormald on size algorithms we prove here that Achlioptas processes cannot postpone the appearance of the giant for more than $0.9455n$ steps whp.

\end{abstract}

\section{Introduction}

We consider here on-line algorithms for Achlioptas processes, introduced by Dimitris Achlioptas. Given a initially empty graph $G$ on $n$ vertices, we launch a random process that at each step selects independently and uniformly at random two edges from the set of non-edges. We must choose one of the two edges and add it to the graph while discarding the other. The goal is to avoid the appearance of a connected component spanning $\Omega(n)$ vertices for as many steps as possible. In the literature these components are referred to as \emph{giant} components. A related class of processes is the class of off-line Achlioptas processes having the same setup except that we are allowed to see the whole sequence of pairs of edges at once, rather than in steps. 

The question Dimitris Achlioptas initially posed was: is there any strategy to choose in between the two edges that can avoid creating a giant for at least $cn$ steps for a constant $c>\frac{1}{2}$ whp? While the answer to Achlioptas's question is affirmative (see \cite{BohFri01} and \cite{SpeWor07}) it is also known that there is a limit on how much on-line Achlioptas processes can postpone the giant's appearance. Even though an optimal strategy was not yet determined the currently known upper bound is $0.96445n$ steps.

The purpose of this work is to determine a new upper bound for on-line Achlioptas processes. We have previously tackled the same problem in \cite{Cob13}, but the results there were missing a final rigorous counting argument, which was there replaced by an empirical observation.

\subsection{Related work}

In \cite{BohFri01} Bohman and Frieze showed that the power of choice in between the two edges can delay the giant for at least $0.535n$ steps whp. In the same paper, they also proved that no algorithm (in the on-line or off-line setup) can postpone the appearance of the giant for more then $n+o(n)$ steps whp. They used an edge density argument. Later in \cite{BoFrWo04} Bohman et al. proved that the giant cannot be delayed for more than $t_{A}n=0.97765n$ steps in the off-line setup or more than $t_{B}n=0.96445n$ steps in the on-line setup using again an edge density argument. We will make use of the same type of argument here.

The main idea exploited in \cite{BoFrWo04} by Bohman et al.\ is that the ability to postpone the giant's appearance is limited by the evolution of the number of vertices in small components throughout the process. Here we exploit only the limitations caused by isolated vertices. Roughly speaking, we prove that a giant will occur before the moment when the number of isolated vertices and the number of edges in our graph sum up to significantly more than the total number of vertices. The intuition around this idea can be states as follows: if this inequality holds, then whp there is a linear number of redundant edges, i.e.\ edges that when added to the graph join two vertices that were already in the same connected component. This indicates that the giant component has already appeared. We leave it to reader to prove that before the giant's appearance the number of redundant edges is sublinear whp (or refer to \cite{SpeWor07}).

In \cite{SpeWor07} Spencer and Wormald studied bounded size strategies. Their work relies on the solution of a differential equation system to determine the proportion of vertices in components of size $s$, for all $s \in [1,K]$ with $K$ a constant, and in components strictly larger than $K$ under different size strategies. These results are further used to determine the moment when the first giant component occurs under some of these strategies. We will determine the minimum number of isolated vertices that an on-line algorithm can leave in our graph at any step using the same differential equation technique.

\subsection{The main result}
\label{subsec:The main result}

We combine the methods used in \cite{BoFrWo04} and \cite{SpeWor07} to prove the following result.

\begin{theorem}
\label{theorem:On-line bound}
Each on-line Achlioptas process whp produces graphs that have components of size $\Omega(n)$ before $t_Cn=0.9455n$ steps.
\end{theorem}

The proof is by contradiction. We accomplish this by proving the following two intermediate results.

\begin{theorem}
\label{theorem:The minimum number of isolated vertices}
Let $i(t)$ be the proportion of isolated vertices at step $tn$ in the graph produced by an Achlioptas process. Then whp we have that $i(0.9455)>0.0548$.
\end{theorem}

\begin{theorem}
\label{theorem:The incapacity}
Let $\epsilon > 0$ be a constant and let $G_{tn}$ be the graph produced by an Achlioptas process, at step $tn$. If $G_{tn}$ contains no component of linear size in $n$ then whp the following inequality holds: \begin{eqnarray}
(1+\epsilon)(1-i(t)) &\geq& t. \nonumber
\end{eqnarray} where $i(t)$ denotes the proportion of isolated vertices at step $tn$.
\end{theorem}

The remainder of this paper is organized as follows. In Section~\ref{subsec:The minimum proportion of isolated vertices} we prove Theorem~\ref{theorem:The minimum number of isolated vertices} while Section~\ref{subsec:The edge density argument} contains the proof of Theorem~\ref{theorem:The incapacity}. Finally, we summarize the main idea and express our intuition about the "strength" of this upper bound in Section~\ref{sec:Conclusions}.

\section{The new upper bound}

Let $G$ be the graph produced by our on-line Achlioptas process at round $tn$ and let $H=\{[n],\{e_1,e'_1,...e_{tn},e'_{tn}\}\}$, where $n$ is the number of vertices, $[n]$ is the set of vertices and $e_i$ and $e'_i$ are the edges presented at round $i$. Also we say that a round is a $(a,b,c,d)$ round when the involved components (i.e.\ the components joined by the two presented edges) have sizes $a,b,c$ and $d$, respectively.

The following density lemma is used as a part of the argument in \cite{BoFrWo04} regarding the upper bounds of on-line and off-line algorithms.

\begin{lemma}
\label{lemma:The Density}
For constants $\epsilon,\delta>0$  and $m=tn \leq n$ let $A_{\epsilon,\delta}$ be the event that there exists $S\subseteq[n]$ such that $|S|<\delta n$ and the subgraph of $H$ induced by $S$ contains more than $(1+\epsilon)|S|$ edges. If $\delta=\delta(\epsilon)=2\epsilon(4te)^{-1-1/\epsilon}$ then $Pr(A_{\epsilon,\delta})=o(1)$.
\end{lemma}

We do not prove this lemma here as we will use it as it is (for a full proof refer to \cite{BoFrWo04}).

\subsection{The minimum proportion of isolated vertices}
\label{subsec:The minimum proportion of isolated vertices}

In this section we prove Theorem~\ref{theorem:The minimum number of isolated vertices} by solving a differential equation system. We use some of the results and techniques in \cite{SpeWor07}.

Now consider the on-line Achlioptas process strategy which is greedy with respect to the number of isolated vertices (i.e.\ at each round its decision minimizes the number of isolated vertices in the graph), which makes no difference in between components of sizes greater than 1 and that takes the first edge in case of a tie after applying the first two criteria. In other words, if we let $w:\{1,2,...,n\}\rightarrow\{1,2\}$ with $w(1)=1$ and $w(s)=2$ for every $s>1$, then, at a $(a,b,c,d)$ round, we choose the first edge if \[w(a)+w(b)\leq w(c)+w(d)\] and the second edge otherwise. This is a bounded size algorithm for $K=1$ and it is in fact the $MinP_1$ strategy studied in \cite{SpeWor07}.

Considering the limitations imposed by the on-line setup (i.e.\ the inability to see the pairs of edges at the next rounds), among the on-line algorithms, this strategy is also optimal with respect to the number of isolated vertices. Thus, there is no on-line algorithm that can leave less isolated vertices (in expectation) than $MinP_1$ does up until any round $tn \leq n$.

Now let $i(G_{tn})$ denote the proportion of isolated vertices in our graph after $tn$ steps under the $MinP_1$ strategy. It has been proven in \cite{SpeWor07} that in the bounded size algorithm setup the proportion of isolated vertices is concentrated whp around the values of a function $i(t)$, for $t\in[0,1]$, and that this function can be determined as the solution of a differential equation system. These results are in fact corollaries of Theorem 1.1, Theorem 2.1 and Theorem 4.1 in \cite{SpeWor07}. The reader should know that we used a slightly different setup than the one used in \cite{SpeWor07}, the main difference being that Spencer et al.\ parametrized $\frac{n}{2}$ steps as one time unit while we decided that parametrizing $n$ steps as one unit is more appropriate. The approach in \cite{SpeWor07} is due to the fact that Achlioptas processes are in fact a "game-like variation" of the much more famous Erd\H{o}s-R\'enyi random graph process and the giant's appearance in this later process takes place at roughly round $\frac{n}{2}$.

The next result is a corollary of Theorem 1.1 in \cite{SpeWor07}.

\begin{theorem}[Spencer, Wormald]
\label{theorem:The number of isolated vertices is concentrated}
There exists a function $i:[0,1]\rightarrow[0,1]$ such that whp for all $t\in[0,1]$ we have that $i(G_{tn})=i(t)+o(1)$.
\end{theorem}

Following the approach in \cite{SpeWor07} we find a relationship in between the $i(t)$ and its derivative. As Achlioptas processes are in fact discrete we need to find the discrete analogue of this relationship first and than talk about the asymptotic behavior as $n$ tends to infinity.

We define $\Delta(a,b,c,d)$ to be the change in the number of isolated vertices when presented with a round where the involved components (the components joined by the two presented edges) have sizes $a,b,c$ and $d$ and the round is not redundant (meaning that none of the two edges joins two vertices already in the same component). Also let $x$ denote the proportion of vertices before a round of the form $(a,b,c,d)$. 

We distinguish the following six cases in which $\Delta(a,b,c,d)\neq0$.

\newcounter{casenum}
\newenvironment{caseof}{\setcounter{casenum}{1}}{\vskip.5\baselineskip}
\newcommand{\case}[2]{\vskip.1\baselineskip {\bfseries Case \arabic{casenum}:} #1\\#2\addtocounter{casenum}{1}}
\begin{caseof}
\case{$(a,b,c,d)=(1,1,s_3,s_4)$, with $s_3,s_4 \geq 1.$}
{
	The probability of such a round is $x^2$ and $\Delta(a,b,c,d)=-2$ as $MinP_1$ chooses the first edge.
}
\case{$(a,b,c,d)=(1,s_2,s_3,s_4)$, with $s_2>1$ and $(s_3,s_4)\neq(1,1).$}
{
	The probability of such a round is $x(1-x)(1-x^2)$ and $\Delta(a,b,c,d)=-1$ as $MinP_1$ chooses the first edge.
}
\case{$(a,b,c,d)=(s_1,1,s_3,s_4)$, with $s_1>1$ and $(s_3,s_4)\neq(1,1).$}
{
	The probability of such a round is $x(1-x)(1-x^2)$ and $\Delta(a,b,c,d)=-1$ as $MinP_1$ chooses the first edge.
}
\case{$(a,b,c,d)=(s_1,s_2,1,1)$, with $(s_1,s_2)\neq(1,1).$}
{
	The probability of such a round is $(1-x^2)x^2$ and $\Delta(a,b,c,d)=-2$ as $MinP_1$ chooses the second edge.
}
\case{$(a,b,c,d)=(s_1,s_2,1,s_4)$, with $s_1,s_2,s_4>1.$}
{
	The probability of such a round is $x(1-x)^3$ and $\Delta(a,b,c,d)=-1$ as $MinP_1$ chooses the second edge.
}
\case{$(a,b,c,d)=(s_1,s_2,s_3,1)$, with $s_1,s_2,s_3>1.$}
{
	The probability of such a round is $x(1-x)^3$ and $\Delta(a,b,c,d)=-1$ as $MinP_1$ chooses the second edge.
}
\end{caseof}

So the expected change in the proportion of isolated vertices after a round is \[\Delta=-4x+4x^2-4x^3+2x^4.\] This is our discrete analogue of a derivative. Also note that at time $0$ the proportion of isolated vertices equals $1$.

The next two results are corollaries of Theorem 2.1 and Theorem 4.1 in \cite{SpeWor07}.

\begin{theorem}[Spencer, Wormald]
\label{theorem:The differential equation has a solution}
There exists $i:[0,1]\rightarrow[0,1]$, a unique solution to the differential equation \[ i'(t)=-4i(t)+4i^2(t)-4i^3(t)+2i^4(t)\] with initial conditions \[i(0)=1.\]
\end{theorem}

\begin{theorem}[Spencer, Wormald]
\label{theorem:The number of isolated vertices is concentrated around the solution}
Let $i$ be the solution to the differential equation in the previous theorem. Then whp \[\lvert i(G_{tn})-i(t) \rvert=\mathcal{O}(n^{-1/4})\] uniformly for $0\leq t \leq1$.
\end{theorem}

The differential equation system in Theorem~\ref{theorem:The differential equation has a solution} can be solved using numerical methods. In this case we choose explicit Runge-Kutta method, just like in \cite{SpeWor07}. The result is a strictly decreasing function with $i(0.9455)\in[0.0548,0.0551]$. We mention that the global truncation error caused by the approximation method we used is smaller than $0.0001$.

\textbf{Note.} The reader can quickly test this numerical results using the Mathemathica code in Appendix~\ref{chap:The Mathematica code snippet}.

\subsection{The edge density argument}
\label{subsec:The edge density argument}

In this section we prove Theorem~\ref{theorem:The incapacity}. After that Theorem~\ref{theorem:On-line bound} follows immediately.

Suppose that $G$ has no component larger than $\delta n$ and that $A_{\epsilon,\delta}$ does not hold (as defined in the setup of Lemma~\ref{lemma:The Density}). We then apply the falseness of $A_{\epsilon,\delta}$ to the components of the subgraph of $G$ induced by $T=[n]\setminus V_1$, where $V_1$ denotes the set of isolated vertices in $G$. We obtain \begin{eqnarray} (1+\epsilon)|T| &\geq& |E(G[T])|.\nonumber \end{eqnarray} which can be rephrased as  \begin{eqnarray} (1+\epsilon)(1-i(t)) &\geq& t. \label{eq:The incapacity} \end{eqnarray} This finishes the proof of Theorem~\ref{theorem:The incapacity}.

But inequation~\ref{eq:The incapacity} cannot hold for small enough $\epsilon$ and for $t=0.9455$ since we have $i(0.9455)>0.0548$, no matter what on-line strategy we use. We conclude that no on-line algorithm can postpone the appearance of the giant for more than $t_Cn=0.9455n$ rounds.

\section{Conclusions}
\label{sec:Conclusions}

So no Achlioptas process can postpone the appearance of the giant for more than $0.9455n$ rounds in the on-line setup. Additionally, we are confident that the same edge density argument corroborated with the differential equation method for random graphs can be used further to get an even smaller value of the upper bound. That is because the evolution of the number of small components throughout the process is known (from \cite{BoFrWo04}) to put a limit on the performance of Achlioptas processes. Informally, almost all the small components are trees, thus accommodating one edge less than vertices. This phenomenon will concentrate edges outside these small components and a giant will occur in the rest of the graph at some moment. 

Here we studied the $MinP_1$ algorithm, a strategy that minimizes the proportion of isolated vertices, i.e.\ components of size 1. But one could study a strategy that minimizes the number of components smaller or equal in size to a constant $K>1$ or even of size sublinear in $n$, probably via a much more involved argument. Empirical results in this direction (that also led to this work) suggest an upper bound as low as $0.9n$ steps.

\appendix
\section{The Mathematica code snippet}
\label{chap:The Mathematica code snippet}

The following Mathematica code snippet can be used to numerically solve the differential equation system that determines the proportion of isolated vertices under the $MinP_1$ strategy, denoted here by $f$. We use the explicit Runge-Kutta method. The results are finally plotted. This code runs under Mathematica 9.0.1.

\begin{lstlisting}[language=Mathematica,
	basicstyle=\footnotesize,
	breakatwhitespace=true,
	breaklines=true,
	frame=l,
  	xleftmargin=\parindent,
  	aboveskip=3mm,
  	belowskip=3mm,
  	captionpos=b,
  	caption=The Mathematica 9.0.1 code snippet for the differential equation system that determines the proportion of isolated vertices under the $MinP_1$ strategy at each round.]
Needs["DifferentialEquations`NDSolveProblems`"];
Needs["DifferentialEquations`NDSolveUtilities`"];
sol=NDSolve[{f'[x]==-(4f[x]-4f[x]^2+4f[x]^3-2f[x]^4),f[0]==1},f[x], {x,0,1},Method->"ExplicitRungeKutta","StartingStepSize"->1/10000];
g[x_]=1-x;
Plot[{Evaluate[f[x]/.sol],g[x]},{x,0,1},PlotRange->All]
\end{lstlisting}

\bibliographystyle{plain}
\bibliography{refs}

\end{document}